\documentclass[reprint,superscriptaddress,nofootinbib,amsmath,amssymb,aps,pra,floatfix]{revtex4-1}
\usepackage{caption}
\usepackage{graphicx}
\usepackage{dcolumn}
\usepackage[inline]{enumitem}
\usepackage[hidelinks]{hyperref}

\usepackage{bm}
\usepackage{amsmath}
\usepackage[font=small]{caption}
\usepackage[labelformat=empty, position=top]{subcaption}
\usepackage[export]{adjustbox}
\usepackage{blindtext}
\usepackage{amssymb}
\usepackage{newunicodechar}
\usepackage{nameref}

\usepackage{lipsum}

\newcommand\copynote[1]{%
  \begingroup
  \renewcommand\thefootnote{}\footnote{#1}%
  \addtocounter{footnote}{-1}%
  \endgroup
}

\begin{document}

\title{Bispectrum-based Cross-frequency Functional Connectivity: Classification of Alzheimer's disease}

\author{Dominik Klepl}
  \affiliation{Centre for Computational Science and Mathematical Modelling, Coventry University, Coventry CV1 2JH, UK 
  }

\author{Fei He}
  \email{fei.he@coventry.ac.uk, klepld@uni.coventry.ac.uk}
  \affiliation{Centre for Computational Science and Mathematical Modelling, Coventry University, Coventry CV1 2JH, UK
}

\author{Min Wu}
  \affiliation{Institute for Infocomm Research, Agency for Science, Technology and Research (A*STAR), 138632, Singapore
}

\author{Daniel J. Blackburn}
  \affiliation{Department of Neuroscience, University of Sheffield, Sheffield, S10 2HQ, UK}
  
\author{Ptolemaios G. Sarrigiannis}
  \affiliation{Department of Neurophysiology, Royal Devon and Exeter NHS Foundation Trust, Exeter, EX2 5DW, UK}


\begin{abstract}
Alzheimer’s disease (AD) is a neurodegenerative disease known to affect brain functional connectivity (FC). Linear FC measures have been applied to study the differences in AD by splitting neurophysiological signals such as electroencephalography (EEG) recordings into discrete frequency bands and analysing them in isolation. We address this limitation by quantifying cross-frequency FC in addition to the traditional within-band approach. Cross-bispectrum, a higher-order spectral analysis, is used to measure the nonlinear FC and is compared with the cross-spectrum, which only measures the linear FC within bands. Each frequency coupling is then used to construct an FC network, which is in turn vectorised and used to train a classifier. We show that fusing features from networks improves classification accuracy. Although both within-frequency and cross-frequency networks can be used to predict AD with high accuracy, our results show that bispectrum-based FC outperforms cross-spectrum suggesting an important role of cross-frequency FC.
\newline

\indent \textit{Clinical relevance}— This establishes diagnostic relevance of cross-frequency coupling in Alzheimer’s disease.
\end{abstract}

\maketitle

\section{Introduction}
Alzheimer's disease (AD) causes early neural degradation leading to cell death and synaptic loss, and is the most common form of dementia. 
Studies have shown that AD alters functional connectivity (FC) over multiple scales \cite{pievani2011functional, konig2005decreased, jeong2004eeg}. Electroencephalography (EEG) is a commonly used method to study and classify AD. The main EEG characteristics associated with AD are slowing of signals and decrease in synchronisation \cite{jeong2004eeg, ghorbanian2015slowing, konig2005decreased,babiloni2016synchronisationreview}. Slowing of EEG in AD was observed as an increased activity in $\delta$ and $\theta$ frequency bands and decreased activity in $\alpha$, $\beta$ and $\gamma$ frequency bands \cite{jeong2004eeg, ghorbanian2015slowing}. Similarly, AD shows changes in synchronisation within low-frequency bands ($<$12 Hz) and is associated with altered FC \cite{babiloni2016synchronisationreview}. However, these characteristics are typically measured on a single EEG or pairwise channel level. In contrast, network based FC analysis can reveal additional characteristics of AD across multiple channels and scales \cite{kabbara2018integrationsegregation, dai2019disrupted}. However, these studies focus only on specific frequency bands, i.e. the within-frequency coupling (WFC). WFC networks of AD have been analysed using various FC methods such as coherence \cite{ blinowska2017classification_coherence}, wavelet coherence \cite{jeong2016waveletcoherence} and phase lag index (PLI) \cite{nobukawa2020classification_PLI}. WFC networks were also used for classification of AD using a multivariate autoregressive model \cite{blinowska2017classification_coherence} and support vector machine (SVM) \cite{nobukawa2020classification_PLI}. 
\copynote{\textcopyright 2022 IEEE.  Personal use of this material is permitted. Permission from IEEE must be obtained for all other uses, in any current or future media, including reprinting/republishing this material for advertising or promotional purposes, creating new collective works, for resale or redistribution to servers or lists, or reuse of any copyrighted component of this work in other works.}

\begin{figure*}[tb]
    \vspace*{2mm}
    \centering
    \includegraphics[width=0.9\linewidth]{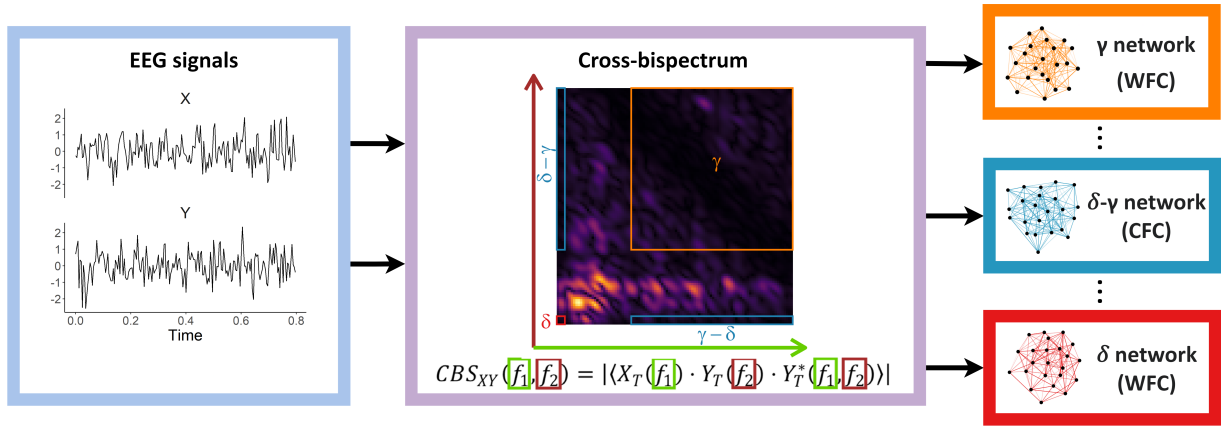}
    \caption{A conceptual schematic of constructing the proposed cross-bispectrum (CBS) networks. First, each EEG signal is cleaned and scaled (left). For each pair of EEG electrodes, CBS is estimated (middle). The frequency bands coupling edge weights are given by the maximum value within the respective CBS window. From each CBS, 25 edges are inferred. This way, 5 within-frequency (WFC) and 20 cross-frequency (CFC) networks are constructed (right).
    }
    \label{overview}
\end{figure*}

This study aims to extend the WFC FC analysis by taking the cross-frequency coupling (CFC) \cite{jirsa2013crossbispectrumcoupling} into account and investigate the predictive power of CFC for AD diagnosis. Only one CFC measure, i.e. phase synchronisation index (PSI), had been used to reconstruct CFC networks of AD \cite{cai2018ADcrossphase}. However, this study did not use any classification methods. Moreover, PSI measures only phase-phase CFC.

(Cross-) Bispectrum (CBS), a higher-order spectral analysis, quantifies quadratic coupling between two frequency components and their algebraic sum \cite{HE2021}. It has been shown to detect phase-amplitude and phase-phase coupling \cite{jirsa2013crossbispectrumcoupling}. CBS-based EEG features have been proposed as biomarkers of epilepsy \cite{mahmoodian2019epilepsybispectrum}, Parkinson's disease \cite{yuvaraj2018PDindexbispectra} and AD \cite{wang2015multiple}. However, these applications study only a few EEG channels. In contrast, our study computes CBS between all pairs of EEG channels to construct both WFC and CFC FC networks (Fig. \ref{overview}).

The CBS estimates of FC are computed to investigate the differences between AD and healthy controls (HC) in nonlinear WFC and CFC, compared to linear WFC measured with cross-spectrum (CS). To the best of our knowledge, this is the first application of CBS in reconstructing FC networks of AD. Finally, we use the estimated networks to train classifiers, in order to examine the predictive power of WFC and CFC.

\section{Data}
EEG recordings were collected from 20 AD patients and 20 HC under 70. A detailed description of the experimental design and confirmation of the diagnosis is provided in \cite{blackburn2018synchronisation}. AD patients that were in the mild to moderate stage with the average Mini Mental State Examination score of 20.1 $(sd = 4)$ were recruited in the Sheffield Teaching Hospital. Age and gender-matched HC participants were recruited. A consent was obtained from all participants.

EEG was acquired using an XLTEK 128-channel headbox, Ag/AgCL electrodes with a sampling frequency of 2 kHz using a modified 10-10 overlapping and 10-20 international electrode placement system with a referential montage with a linked earlobe reference \cite{blackburn2018synchronisation}. 30 minutes long sessions at rest were recorded, consisting of alternating two-minute-long epochs during which the participants had their eyes closed (EC) or open. Only the EC epochs are used in this study.

All the recordings were reviewed by a neurophysiologist on the XLTEK review station with time-locked video recordings. Three 12-second-long artefact-free epochs per subject were isolated. Finally, 23 bipolar channels were created: F8–F4, F7–F3, F4–C4, F3–C3, F4–FZ, FZ–CZ, F3–FZ, T4–C4, T3–C3, C4–CZ, C3–CZ, CZ–PZ, C4–P4, C3–P3, T4–T6, T3–T5, P4–PZ, P3–PZ, T6–O2, T5–O1, P4–O2, P3–O1 and O1–O2 \cite{blackburn2018synchronisation}.

\subsection{EEG pre-processing}
EEG signals were band-pass filtered to 0.1-100 Hz range using a zero-phase 5\textsuperscript{th} order Butterworth filter. 50 Hz power line noise was removed using a zero-phase 4\textsuperscript{th} order Butterworth stop-band filter; and the data were down-sampled to 250 Hz using an 8\textsuperscript{th} order Chebyshev type I filter. The signals were normalised to zero mean and unit standard deviation.

\begin{figure}[tb]
    \centering
    \includegraphics[width=0.92\linewidth]{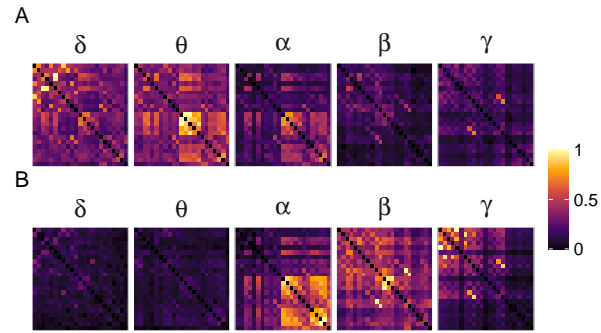}
    \caption{Average connectivity matrices measured with cross-spectrum of (A) AD and (B) HC. For visualisation, the values were min-max normalised.
    }
    \label{connectivity-mat-spec}
\end{figure}

\begin{figure*}[tb]
    \vspace*{1mm}
    \centering
    \includegraphics[width=0.95\linewidth]{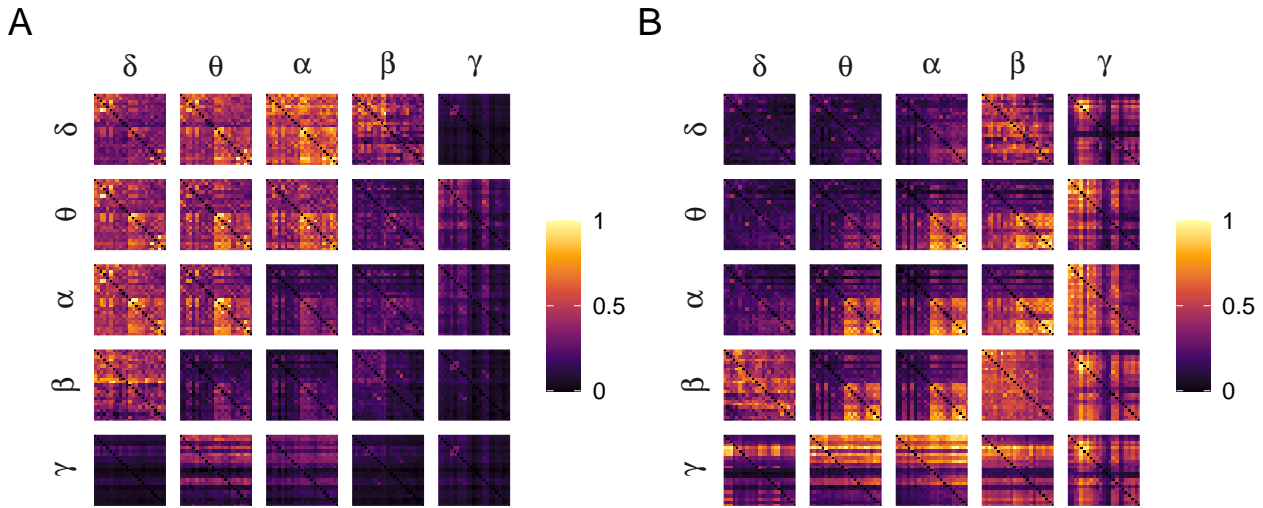}
    \caption{Average connectivity matrices measured with cross-bispectrum of (A) AD and (B) HC with input frequency on the vertical facets and output frequency on horizontal. For visualisation purposes, the values were min-max normalised.
    }
    \label{connectivity-mat-bispec}
\end{figure*}

\section{Methods}

\subsection{Cross-spectrum and cross-bispectrum}
The CS of signal $S_X$ is calculated via smoothed periodogram. Fast Fourier Transform (FFT) is used to estimate the periodogram with Daniell smoothers. The periodogram is computed over 256 frequency bins (0.98 Hz bandwidth). CS at frequency $f$ is then computed as: 
\begin{equation}
CS_{XY}(f) = S_X(f) \cdot S_Y(f).
\end{equation}

A direct FFT-based method is used to estimate CBS:
\begin{equation}
    CBS_{XY}(f_1, f_2) = \langle X_T(f_1) \cdot Y_T(f_2) \cdot Y_{T}^{\ast}(f_1+f_2) \rangle,
\end{equation}
where $\langle \cdot \rangle$ denotes averaging, $X_T(f)$ is a Fourier Transform of signal $X$ over an interval $T$ and $Y^{\ast}$ is the complex conjugate. 256-point FFT is used. CBS is computed over 50 segments with 50\% overlap. The estimated CBS is then smoothed in frequency using a Rao-Gabr window (size 5).

CS and CBS were computed for all pairs of EEG channels. Five frequency bands $b$ are considered: $\delta$ $(0.5-5 Hz)$, $\theta$ $(5-8 Hz)$, $\alpha$ $(8-16 Hz)$, $\beta$ $(16-32 Hz)$ and $\gamma$ $(32-100 Hz)$.

The connectivity $CN$ between channels $X$ and $Y$ and frequency bands $b_X$ and $b_Y$ is computed as:
\begin{equation}
CN_{XY}^{CS}(b_X) = \max(|CS_{XY}(f \in b_X)|), b_X=b_Y,
\end{equation}
\begin{equation}
    CN_{XY}^{CBS}(b_X,b_Y) = \max(|CBS_{XY}(f_1 \in b_X,f_2 \in b_Y)|),
\end{equation}
for CS and CBS respectively. This results in 5 WFC and 20 CFC measures for each pair of channels. Note that CBS edges are directed, thus we collapse them to undirected by taking the average of edge weights.

To reduce the risk of false positives induced by spurious random coupling, surrogate thresholding was used \cite{theiler1992surrogate}. For each pair of channels, 200 surrogate signals were generated using the Fourier transform surrogate method, which scrambles the phase, and their CS and CBS are computed. The 95\% confidence interval of surrogate values is used as a threshold. Coupling values below the threshold are set to zero. We obtain a set of brain networks for each EEG epoch, i.e. $N \times N$ matrices ($N=23$). For CS and CBS, there are 5 and 25 connectivity matrices, respectively.

To further reduce the risk of false-negative edges, the weighted networks are filtered using relative quantile-based thresholding which preserves only the top-k \% of the strongest edges ($k \in \{10, 20, 30, 40, 100\}$.

\subsection{Network classification}
Finally, we use the networks to elucidate their discriminatory power for classification of AD. The upper triangular of adjacency matrix of each network is vectorised and forms a feature vector used to train a classifier. SVM with radial basis kernel is used and principal component analysis is employed, such that the components capture 90\% variance of the data. Moreover, the features are scaled to zero mean and unit standard deviation. A leave-one-subject-out cross-validation is used in this work.

Two categories of classifiers are trained: network level and concatenated-networks level. Network level classifiers are fitted for each constructed network and each network filter to test which frequency coupling types can be used for classification, i.e. show the largest differences between AD and HC. Thus, we fit 125 and 25 (frequency band combinations $\times$ network filters) classifiers using CBS and CS respectively. 

Concatenated-networks level classifiers are fitted by concatenating the vectorised adjacency matrices for each network filter separately. Thus, the feature vectors consists of concatenated adjacency matrices of 25 and 5 networks for CS and CBS, respectively. We hypothesise that the information captured by each network is at least partially unique, thus a classifier trained on the concatenated networks should outperform the classifiers trained on individual networks, as it can leverage the information from all networks.

\section{Results and Discussion}

\begin{figure*}[tb]
    \vspace*{1.2mm}
    \centering
    \includegraphics[width=0.8\textwidth]{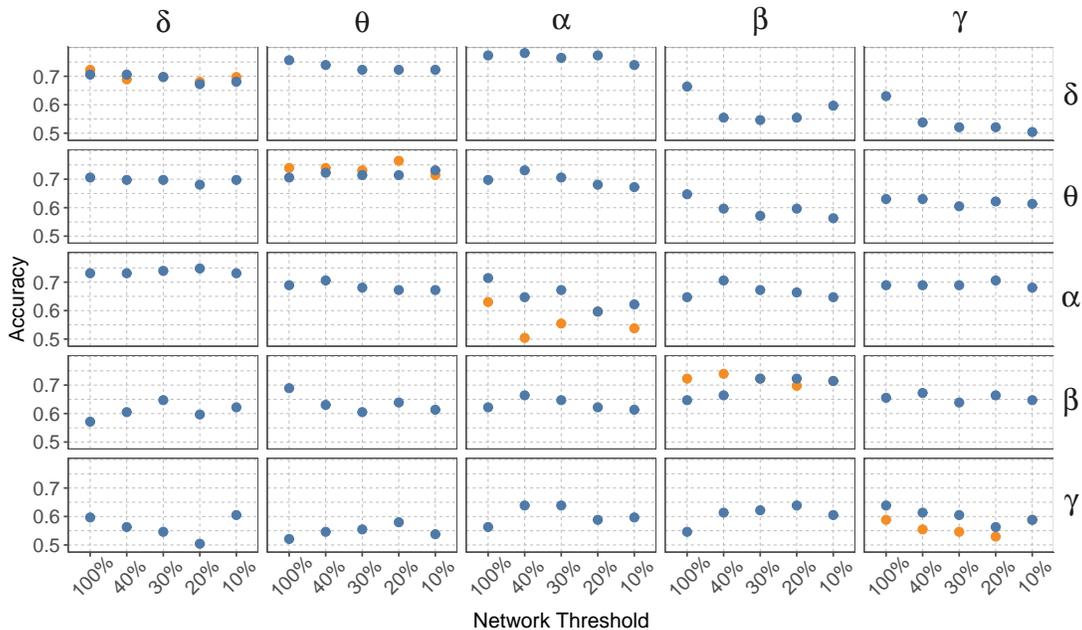}
    \caption{Accuracy of classifiers trained across frequency couplings and filtered networks using cross-spectrum (orange) and cross-bispectrum (blue) networks.}
    \label{bandwise-SVM}
\end{figure*}

\subsection{Connectivity matrices}
Averaged connectivity matrices (Fig. \ref{connectivity-mat-spec} and \ref{connectivity-mat-bispec}) indicate that both CS and CBS seem to detect the differences in the networks. In CS, these differences are most pronounced in $\delta$ and $\theta$ bands where AD have higher FC, and in $\beta$ where AD have lower FC. In CBS, differences can be observed in almost all frequency bands and their couplings. The differences are rather obvious in $\delta$, $\theta$, $\alpha$ and their CFC. Differences in these bands (both WFC and CFC) were reported by Cai et al. \cite{cai2018ADcrossphase} using PSI, but in the opposite direction to our findings. This might be explained by the differences between CBS and PSI, as PSI is a purely phase measure unlike the CBS \cite{jirsa2013crossbispectrumcoupling}. Our findings are consistent with the literature reporting the increased activity in $\delta$ and $\theta$ in AD \cite{jeong2004eeg,konig2005decreased}. Also, the distortion of structure observed in multiple CS and CBS networks is consistent with the disrupted information processing in AD \cite{dai2019disrupted}.

\subsection{Classification results}
First, SVM classifiers were trained for each network separately to compare the discriminatory power of CS and CBS based networks (Fig. \ref{bandwise-SVM}). However, a direct comparison of CS and CBS is only possible for the WFC classifiers (along the diagonal of Fig. \ref{bandwise-SVM}). CS-based models perform better in $\delta$ and $\theta$. On the other hand, CBS-based models perform better in $\alpha$, $\beta$ and in $\gamma$ (100\%). Thus, in the WFC comparison, neither of the methods performs consistently better, suggesting that CBS-based networks are equally capable of capturing the differences in WFC as the CS-based networks.

Additionally, the results show that multiple types of CBS-based CFC networks can deliver comparable or even slightly better accuracy (off diagonal of Fig. \ref{bandwise-SVM}). This is most clear in the $\delta-\theta$, $\delta-\alpha$, $\theta-\delta$, $\theta-\alpha$, $\alpha-\delta$ and $\alpha-\theta$ models. This highlights the importance of studying CFC networks in addition to the WFC networks. 
Next, we trained concatenated-networks models to examine whether leveraging information from multiple networks improves the classification performance. This is confirmed in Table \ref{table: concatenated models}. All CBS-based models outperform the CS-based models suggesting that including information about nonlinear and CFC coupling can be crucial for modelling and classification of AD.

\begin{table}[tb]
\centering
\caption{Performance of concatenated-networks models} 
\begin{tabular}{|l|l|l|l|}
  \hline
  Model & Accuracy & Sensitivity & Specificity \\ 
  \hline
  Spectrum - Top 100\% & 73\% & 71.8\% & 74.2\% \\
  Spectrum - Top 40\% & 72.4\% & 73.7\% & 71.4\% \\ 
  Spectrum - Top 30\% & 72.4\% & 74.3\% & 71\% \\ 
  Spectrum - Top 20\% & 71.3\% & 72.4\% & 70.4\% \\ 
  Spectrum - Top 10\% & 70.7\% & 80\% & 66.4\% \\
  Bispectrum - Top 100\% & 74.1\% & 72.9\% & 75.3\% \\
  Bispectrum - Top 40\% & 73.6\% & 72.6\% & 74.4\% \\ 
  Bispectrum - Top 30\% & 75.9\% & 72.3\% & 80\% \\ 
  Bispectrum - Top 20\% & 77.6\% & 74.2\% & 81.5\% \\ 
  Bispectrum - Top 10\% & 77\% & 81.4\% & 74\% \\ 
   \hline
\end{tabular}
\label{table: concatenated models}
\end{table}

\subsection{Conclusions and Future work}
We have demonstrated that CBS and CS detect similar differences between AD and HC networks, but CBS has an advantage of capturing cross-frequency and nonlinear interactions. Although CBS was shown to be a powerful tool to detect various types of WFC and CFC, such as phase-phase and phase-amplitude, it is not possible to distinguish between them. Therefore, a combination of CBS with other types of CFC methods could be a plausible direction for future research.

Finally, we use relatively simple features for the classification purpose. The use of feature selection methods might further improve the classification performance by removing uninformative and correlated features. In future research, it might also be important and interesting to explore graph-based features that would capture the differences between AD and HC in a lower-dimensional space more efficiently.

\section*{Acknowledgement}
The original data collection was funded by a grant from the Alzheimer’s Research UK (ARUK-PPG20114B-25). The views expressed are those of the author(s) and not necessarily those of the NHS, the NIHR or the Department of Health.

\newpage
\bibliographystyle{unsrt}
\bibliography{ref}

\end{document}